\documentclass[aps,twocolumn,showpacs,floatfix]{revtex4}

\usepackage{mathrsfs}
\usepackage{dcolumn}
\usepackage{bm}
\usepackage{amsmath}
\usepackage{amsthm}
\usepackage{amssymb}
\usepackage{graphicx}
\usepackage{footmisc}
\usepackage{nicefrac}
\usepackage{appendix}

\newcommand{\eref}[1]{(\ref{#1})}

\usepackage{epstopdf}

\begin{document}

\title{Resonance scattering and the passage to bound states in the field of near-black-hole objects}
\author{G. H. Gossel}
\affiliation{School of Physics, University of New South Wales, Sydney 2052,
Australia}
\author{V. V. Flambaum}
\affiliation{School of Physics, University of New South Wales, Sydney 2052,
Australia}
\author{J. C. Berengut}
\affiliation{School of Physics, University of New South Wales, Sydney 2052,
Australia}
\date{\today}

\begin{abstract}
We examine the spectrum of a massive scalar particle interacting with the strong gravitational field of a static, spherically symmetric object which is not quite massive enough to be a black hole. As was found in the case of massless particles, there exists a dense spectrum of long lived resonances (meta-stable states), which leads to an energy-averaged cross section for particle capture which approaches the absorption cross-section for a Schwarzschild black hole. However, the generalisation to non-zero mass introduces new phenomena, along with important qualitative changes to the scattering properties. In contrast to the massless case, there exists a spectrum of bound states with almost identical structure to that of the resonances, allowing for the possibility of radiative transitions and particle capture. The resonance lifetimes for elastic processes are parametrically larger than for massless particles, meaning the absorption cross-section approaches the black hole case faster than for massless scalars.
\end{abstract}


\maketitle  

\section{Introduction}
\label{sec:Introduction}
In this work we analyse the scattering properties of massive scalar particles in strong gravitational fields. We find the resulting dynamics to be significantly richer and more complex than in the massless scalar case we considered previously~\cite{ScalarRes}. The strong gravitational field is provided by a near-black-hole object, defined as having a radius $R$ that slightly exceeds the corresponding Schwarzschild radius $r_s = 2GM/c^2$, and taken to be static and spherically symmetric. As with the massless particle case, we find the spectrum to be dominated by a set of dense, narrow resonances which give rise to black-hole-like absorption as $r_s\to R$ (black hole limit). That is, the resonances lifetimes are such that an observer will not live long enough to get the scattered particle back even for elastic processes. However, this system also possesses bound states, the energies of which are given by the same formula used to describe the resonance positions. Having both bound and resonant states, this system may have radiative transitions between resonant and bound states, which drive inelastic capture. Furthermore, we find that the lifetime (for elastic processes) of the resonances for massive incident particles is parametrically larger than the massless case. We show that this increase in lifetime means the system acquires black-hole-like absorption well before an equivalent system with massless incident particles.

In previous works we found that for massless spin-$0$, $1/2$, and $1$ particles scattering in the field of a near-black-hole object, there exists a dense spectrum of narrow resonances (long lived meta-stable states), the lifetimes and density of which tend to infinity as the body $r_s\rightarrow R$ \cite{ScalarRes,DiracRes,BosonRes}. The resulting energy average cross-section for capture into these resonances (optical cross-section) exactly matched that of a Schwarzschild black hole in the low energy limit. This illustrated that the body may develop black hole absorption properties before the formation of a black hole. On the other hand, this set of resonances is complemented by an equally dense set of bound states for massive particles, the spectrum of which collapses to zero energy (and becomes quasi-continuous) in the black hole limit \cite{Gossel,DiracBound}. The requirement that the metric be static and spherically symmetric does not in turn require isotropy of the stress energy tensor, freeing us to explore the parameter space beyond the well known Buchdahl limit $r_s/R<8/9$. The ability to take $r_s/R\to1$ for anisotropic spheres has been examined previously (see e.g.~\cite{Anisotropy1,Anisotropy2,Anisotropy3,Anisotropy4}). Two examples of such metrics were developed by Florides~\cite{Flor74} and Soffel~\cite{Soffel77}, and are discussed further in later sections.

In this work we consider the effect of non-zero particle mass on the scattering of scalar particles around finite-size bodies. We show that a dense spectrum of resonances still exists for massive particles, but with two important differences. Firstly, the width of a given resonance is smaller for a particle with mass $m\neq0$ and thus the lifetime of the state is longer (with the resonance position being weakly dependent on $m$). This longer lifetime implies that the transition to black-hole-like absorption occurs much earlier than in the massless case.

Secondly, there exists a spectrum of bound states for $\epsilon<m$, `shifting' the start of the resonance spectrum to higher energies $\epsilon>m$. Interestingly, the structure of the resonance and bound level energies is nearly identical for $r_s$ close to $R$: the energy levels of both bound and unbound states may be described by the same formula. Therefore, as $r_s\to R$ (with fixed particle mass $m$) any resonance of principal quantum number $n$ (defined as a resonance having $n-1$ nodes on the interior, consistent with the bound spectrum) will disappear from the unbound spectrum and be replaced by a bound state. 

In the unbound case, as $r_s \to R$, all resonances of a given $n$ will eventually cross into the bound state spectrum. However, for a given energy range $\Delta \epsilon$ in the continuum, the resonance energy spacing $D$ and width $\Gamma_n$ tend to zero (lifetime $\tau \rightarrow \infty$), while their ratio remains finite. This allows us to define the total cross-section for particle capture into these long-lived states using the optical model~\cite{LLV3}, which is calculated by averaging over a small energy interval containing many resonances. In the low momentum limit the resonance capture cross-section is $\sigma_{a}=4\pi^2 r_{s}^{3}\epsilon^3/p^2$ where $\epsilon$ is the energy of the incident particle and $p$ is the particle momentum at infinity. This is precisely the low momentum absorption cross section for massive scalar particles incident upon a Schwarzschild black hole \cite{Unruh}.

Additionally, such a system may have radiative transitions allowing particles to drop down from resonances into bound states. While we only calculate the resonance spectrum for s-wave ($l=0$) states, all states including those with $l\neq0$ contribute to the total inelastic width that appears in the formula for the cross section. We show that so long as the individual partial widths of any inelastic processes are negligible compared to the distance between the levels, only the elastic width contributes to the total energy-averaged capture cross-section given above.

It should be noted that the resonances considered in this work are localised (in the black hole limit) entirely on the interior of the finite-size body, and thus differ from the quasi-bound states that exist in the exterior black hole spacetime (see e.g. \cite{KerrElectron,Gaina,Grain,Lasenby2,Pindzola,Barranco,Pravica,Glampedakis}).

As in previous works we perform both analytical and numerical calculations, with good agreement between the two.
\section{Wave equations}
\label{sec:GeneralCase}
The Klein-Gordon equation for a scalar particle of mass $m$ in curved
space-time with the metric $g_{\mu \nu}$ ($c~=~\hbar~=1$) is
\begin{equation}
\label{eq:KG}
\partial_\mu (\sqrt{-g} g^{\mu\nu}\partial_\nu \Psi)+\sqrt{-g} m^2\Psi =0.
\end{equation}
One may write the metric of a static, spherically symmetric non-rotating body as 
\begin{equation}
\label{eq:GenMetric}
ds^{2}=  e^{\nu \left( r \right)}dt^{2} - e^{\lambda \left( r \right)}dr^{2} - r^{2}d\Omega^{2} .
\end{equation}
Substituting Eqn.~\eref{eq:GenMetric} into Eqn.~\eref{eq:KG} and applying the separation of variables $\Psi (x)=e^{-i\epsilon t}\psi (r)Y_{lm}(\theta ,\phi )$ yields the general radial wave equation
\begin{multline}
\label{eq:GenRadialEqn}
\psi''(r)+ \left[\frac{\nu '(r) - \lambda'(r)}{2} + \frac{2}{r}\right]\psi'(r)  \\
+\left[e^{\lambda(r) - \nu(r)}\epsilon ^{2}-e^{\lambda(r)} m^2\right]\psi (r) = 0 ,
\end{multline}
where we have set $l=0$ (s-wave only) for simplicity.
\section{General Interior Solution}
\label{sec:GenIntSol}
Equation~\eref{eq:GenRadialEqn} can be transformed into a Schr\"odinger-like equation by making the substitution $\psi(r)~=~ e^{[\lambda(r)-\nu(r)]/4}\chi(r)/r$, yielding
\begin{equation}
\label{eq:SchrodEqn}
-\chi''(r)+\frac{1}{4}\chi(r)\left[2\alpha'(r)+\alpha^2(r)-4\beta(r) \right]=0,
\end{equation}
where
\begin{gather}
\alpha(r)=\frac{1}{2}\left[\nu'(r)-\lambda'(r)\right]+\frac{2}{r},\notag{} \\
\beta(r) = e^{\lambda(r)}\left(e^{-\nu(r)}\epsilon^2-m^2\right).
\end{gather}
In the limit $r_s\to R$ the metric coefficient $e^{\nu(r)} \to0$ (smoothly) on the interior $r\leq R$ as time slows down in the black hole limit. Thus in this limit only the energy term contributes to the bracketed expression in Eqn.~\eref{eq:SchrodEqn}; the energy term becomes large as the wavefunction oscillates many times in the strong field on the interior allowing us to invoke the semi-classical approximation. Specifically, we may ignore the contributions from $\nu'(r)$ and $\lambda'(r)$ in favour of the metric coefficient exponentials due to the smoothness of $e^{\nu(r)}$, and the general form of $e^{\lambda(r)}$ (see below). As both the function $e^{\nu(r)}$ and its derivative tend to zero as $r_s \to R$, we may write
\begin{align}
\label{eq:nuDeriv}
\left(e^{\nu(r)}\right)' = \nu'(r)e^{\nu(r)} \to 0 \\ \notag{}
\implies \nu'(r) \ll e^{-\nu(r)}.
\end{align}
For spherically symmetric static spheres joined to the Schwarzschild exterior at $r=R$, $e^{\lambda(r)}$ always takes the form \cite{Anisotropy1,TwoSphere}
\begin{equation}
\label{eq:LambdaDefn}
e^{\lambda(r)}=\left(1-\frac{r_s r^2 }{R^3}\right)^{-1}.
\end{equation}
These assumptions serve to define the set of metrics for which our analysis and results are valid, a simple example of which is the Florides metric~\cite{Flor74} discussed in Sec.~\eref{sec:NumericCalcs}.

Using Eqns.~\eref{eq:nuDeriv} and~\eref{eq:LambdaDefn} we see that in the black hole limit $\beta(r)$ dominates over $\alpha(r)$ and its derivatives in Eqn.~\eref{eq:SchrodEqn} everywhere except for a small region near the boundary (where $\lambda'(r)\to \infty$), the size of which tends to zero as $r_s\to R$. For brevity we continue to write $e^{\lambda(r)}$ rather than the explicit form above until a specific $e^{\nu(r)}$ is considered.

Taking only the dominant term in Eqn.~\eref{eq:SchrodEqn} in the limit $r_s\to R$ would result in neglecting the $\alpha(r)$, $\alpha'(r)$, and $m^2$ terms in the effective potential. However, we are interested in the effect of non-zero $m$ and therefore keep it in the semi-classical momentum, now written as
\begin{align}
k(r)&=\sqrt{\beta(r)},\notag{} \\
&=e^{\lambda(r)/2} \left[  e^{-\nu(r)}\epsilon^2-m^2\right]^{1/2}.
\end{align}
This expression appears in the semi-classical interior wave function 
\begin{equation}
\label{eq:IntSol}
\chi(r)=\frac{a}{\sqrt{k(r)}}\sin \left( \int_{0}^{r}k(r')dr'\right),
\end{equation}
where $a$ is a constant pre-factor determined by overall normalisation. This semi-classical solution is valid so long as $\beta(r)$ dominates over $\alpha(r)$ and its derivatives in Eqn.~\eref{eq:SchrodEqn}, as discussed above.

 Transforming back to the original wavefunction $\psi$ yields
\begin{equation}
\psi(r) = a\frac{e^{\nicefrac{-\nu(r)}{4}}}{r \left(\epsilon^2 e^{-\nu(r)}-m^2\right)^{1/4}}\sin \left(\int_{0}^{r}k(r')dr'\right).
\end{equation} 
To simplify the above expression we invoke the black hole limit and neglect $m$ in the amplitude pre-factor, but allow it to remain in the phase integral. This is justified by noting that the narrow resonances we wish to investigate, as discussed in~\cite{ScalarRes}, are far more sensitive to the rapidly varying (in energy) phase at the boundary than they are to the (slowly varying in energy) amplitude pre-factor. In this case the pre-factor simply reduces to $\nicefrac{a}{(r\sqrt{\epsilon})}$. In propagating the interior solution to large distances only the logarithmic derivative of the interior solution is required, thus the constant pre-factor may be neglected, giving the interior solution as
\begin{equation}
\label{eq:IntSolnFinal}
\psi(r) =\frac{1}{r} \sin \left(\int_{0}^{r}k(r')dr'\right).
\end{equation}
It is useful for later calculations to define the total phase accumulated at the boundary as
\begin{align}
\Phi(r_s)&=\int_{0}^{R} k(r) dr,\notag{}\\ 
&=\int_{0}^{R}  \left[ \epsilon^2 e^{\lambda(r)-\nu(r)}-m^2e^{\lambda(r)}\right]^{1/2}dr.
\end{align}
In the black hole limit we expand around $e^{\nu(r)}\ll \epsilon^2/m^2$ (over all $r\leq R$) allowing us to express the integrand as
\begin{equation}
\label{eq:SepIntegrals}
\Phi(r_s) = \epsilon \int_{0}^{R} e^{[\lambda(r)-\nu(r)]/2}dr- \frac{m^2}{2\epsilon}\int_{0}^{R}e^{[\lambda(r)+\nu(r)]/2}dr.
\end{equation}
Defining $x=  \int_{0}^{R} \! e^{[\lambda(r)-\nu(r)]/2} \, dr$, $y=\int_{0}^{R} \!e^{[\lambda(r)+\nu(r)]/2}dr$ allows us to write the interior phase at the boundary as
\begin{equation}
\label{eq:PhaseFinal}
\Phi(r_s) = \epsilon x - \frac{m^2 y}{2\epsilon},
\end{equation}
where both $x$ and $y$ have units of length, and $x\rightarrow \infty$, $y\rightarrow0$ in the black hole limit. 
\section{Matching to the exterior solution}
Outside a spherically symmetric, non-rotating body of mass $M$ and
radius $R$ the metric is given by
\begin{equation}\label{eq:ExteriorMetric}
ds^2 = \left(1-\frac{r_s}{r}\right)dt^2-\left(1-\frac{r_s}{r}\right)^{-1}dr^2 -
r^2 d\Omega^2 ,
\end{equation}
where $r_s = 2GM$ is the Schwarzschild radius of the body and $G$ is the
gravitational constant. Substitution of Eqn.~\eref{eq:ExteriorMetric} into Eqn.~\eref{eq:GenRadialEqn} gives the exterior radial equation as 
\begin{align}
\label{eq:ExteriorWave}
\psi''(r) &+ \left(\frac{1}{r-r_s}+\frac{1}{r}\right)\psi'(r) \\ \notag{}
+&\left(\frac{r^2 \epsilon ^2}{(r-r_s)^2}-\frac{m^2 r}{r-r_s}
\right)\psi (r) = 0.
\end{align}
If the radius of the body $R$ only slightly exceeds $r_s$ the solution just
outside the body is the following linear combination of the incoming and outgoing waves,
\begin{equation}\label{eq:ExtSol}
\psi \sim \exp [-ir_s\epsilon \ln(r-r_s)]+
\mathcal{R}\exp [ir_s\epsilon \ln(r-r_s)],
\end{equation}
where $\mathcal{R}$ is the reflection coefficient. Here we have ignored contributions from terms containing $m$ under the assumption
\begin{equation}
\label{eq:ExtCond1}
m\ll \epsilon \sqrt{r/(r-r_s)}.
\end{equation}
It will be shown that in the black hole limit the bound and resonance energies may be represented by the same function, given by Eqn.~\eref{eq:GenEnergy}. For these energies it may be shown that Eqn.~\eref{eq:ExtCond1} is valid for sufficiently large principal quantum number $n$. Bound states with large $n$ may always be found for $r_s/R$ sufficiently close to $1$.

By matching the logarithmic derivatives of the interior and exterior solutions, given by Eqns.~\eref{eq:IntSolnFinal} and \eref{eq:ExtSol} respectively, on the boundary $r=R$, we find the reflection coefficient to be 
\begin{equation}
\label{eq:RBa}
\mathcal{R} =- e^{2i\left[\Phi(r_s)-\epsilon r_s \ln(R-r_s)\right]}.
\end{equation}
By redefining $x$ (the term in the phase linear in $\epsilon$) in Eqn.~\eref{eq:PhaseFinal} to absorb the logarithmic term in Eqn.~\eref{eq:RBa}, we may re-write $\mathcal{R}$ as 
\begin{equation}
\label{eq:RB}
\mathcal{R} =- e^{2i\Phi(r_s)}.
\end{equation}
In solving for the reflection coefficient we have made use of the relation
\begin{equation}
(R-r_s)k(R)= \epsilon R
\end{equation}
which follows from imposing continuity of the metric \eref{eq:GenMetric} at the boundary $r=R$ and taking $r_s \to R$.
\section{Scattering matrix}
At large distances Eqn.~\eref{eq:ExteriorWave} is Coulomb-like with effective charge $Z=-r_s (\epsilon^2+p^2)/2$ where $p$ is the momentum of the particle at infinity~\cite{commentP}. This allows the solution to be written in terms of outgoing and incoming waves as 
\begin{equation}
\label{eq:Coulomb_scatter}
\psi(r)= \frac{A e^{i z} +B e^{-iz}}{r} ,
\end{equation}
where $z = p r+\varphi \ln(2p r) + \delta^{C} $, $\varphi =r_s \left(\frac{\epsilon^2}{2p}+\frac{p}{2}\right)$ and $\delta^{C}~=~ \arg\left[\Gamma\left( 1 -i \varphi \right)\right]$ is the Coulomb phase shift. Therefore the s-wave scattering matrix \cite{LLV3} is
\begin{equation}
S_0= (-1) \frac{A}{B}e^{2i\delta^{C}} .
\end{equation}
The coefficients $A$ and $B$ are calculated by constructing solutions to Eqn.~\eref{eq:ExteriorWave} in different regions of the exterior. These solutions are then matched under appropriate conditions allowing propagation of the boundary conditions at $r=R$ to large distances where Eqn.~\eref{eq:Coulomb_scatter} is valid. This procedure is detailed in \cite{ScalarScattering}, with the resulting low energy $S$-matrix is given by
\begin{equation} \label{eq:ScatMatrix}
S_0= \frac{1+\mathcal{R}- \epsilon r_{s}^{2} p C^2 (1-\mathcal{R})}
{1+\mathcal{R}+ \epsilon r_{s}^{2} p C^2(1-\mathcal{R})}e^{2i\delta_C},
\end{equation}
where
\begin{equation}
\label{eq:CFactor}
C^2 = \frac{2\pi \varphi}{1-\exp[-2\pi \varphi]} ,
\end{equation}
is the Coulomb factor.
\section{Resonant Capture}
\label{sec:ResSubSec}
Resonant states with complex energy $\epsilon= \epsilon_n -i\Gamma_n/2$ occur where the $S$-matrix \eref{eq:ScatMatrix} has a pole. To find these complex poles we first set the denominator of \eref{eq:ScatMatrix} to zero, solve for $\mathcal{R}$,  and use \eref{eq:RB} (under the condition $ \epsilon r_{s}^{2} p C^2\ll1$), giving 
\begin{align}
\label{eq:ResStart}
2i \Phi(r_s)=2i \left( \epsilon x - \frac{m^2}{2\epsilon} y\right) = 2\epsilon r_{s}^{2} p C^2 +2 \pi n i.
\end{align}
As we are considering narrow resonances and low energies,  $\Gamma_{n} \ll \epsilon_n \ll 1/R$ (where $1/R$ is the energy scale in our system of units), we make the substitution $\epsilon \rightarrow \epsilon_n -\frac{i}{2}\Gamma_n$ only in the linear energy terms of \eref{eq:ResStart}. Specifically, we do not substitute into the term $\epsilon p$ as doing so yields width terms $\propto \Gamma_{n}^{2}$ which are negligible compared to the terms linear in $\Gamma_n$ arising from other energy terms in Eqn.~\eref{eq:ResStart}. Under these conditions, substituting $\epsilon\to \epsilon_n - i\Gamma_n/2$ in Eqn.~\eref{eq:ResStart} yields
\begin{equation}
\left(2i \epsilon_n +\Gamma_n \right) x - \frac{i m^2\left(\epsilon_n+\frac{i}{2}\Gamma_n \right)y}{\epsilon_{n}^{2}} = 2\epsilon_n r_{s}^{2} p C^2  +2 \pi n i.
\end{equation}
Equating real and imaginary components yields the resonance widths
\begin{align}
\label{eq:GenWidth}
\Gamma_n = \frac{2\epsilon_n r_{s}^{2} p C^2}{x+\frac{m^2 y}{2\epsilon_n^2}},
\end{align}
and positions
\begin{align}
\label{eq:GenEnergy}
\epsilon_n = \frac{\pi n+\sqrt{2m^2 x y +\pi^2 n^2}}{2x}.
\end{align} 
Equation~\eref{eq:SepIntegrals} shows that  in the black hole limit $y/x\to 0$. Under this condition the mass term in Eqn.~\eref{eq:GenEnergy} becomes negligible and the result reduces to the energy spectrum for resonances with $m=0$ given in \cite{ScalarRes}: taking non-zero particle mass does does not significantly perturb the resonance energies, and the energy spacings are given by $D = \frac{\pi}{x}$ as in the massless case. 

However, the presence of the $p$ term in the numerator of $\Gamma_n$ means the width \textit{is} strongly dependent on the mass of the scattered particle. It is worth noting that whilst the $p$ term in $\Gamma_n$ implies the width will tend to zero when $m = \epsilon$, $C^2$ is also a function of $p$ and so this limit evaluates to
\begin{equation}
\lim_{p\rightarrow 0} \Gamma_n = \frac{2\pi r_{s}^{3} \epsilon^3}{x+y/2},
\end{equation}
which gives a lower bound on the resonance widths for a given $r_s$ in the region $\epsilon \gtrsim m$. The above expression is calculated assuming that for $p\to0$, $\exp[\pi \epsilon^2 /p]\gg1$ which is only valid for $m\neq 0$. 
\section{Contributions from Elastic and Inelastic Processes}
The Breit-Wigner formula giving the total cross-section for elastic and inelastic scattering (for an isolated resonance) is
\begin{equation}
\sigma_t = \frac{\pi}{p^2}\frac{\Gamma_{\mathrm{el}}\Gamma_{\mathrm{tot}}}{(\epsilon-\epsilon_n)^2+\Gamma_{\mathrm{tot}}^{2}/4}.
\end{equation}
If we choose a region containing many resonances (i.e. a energy interval of size $\Delta \epsilon\gg D$) the energy averaged capture cross-section of the resonances in this region is
\begin{equation}
\label{eq:XSectionSum}
\overline{\sigma} = N \int_{\epsilon_n-\delta \epsilon/2}^{\epsilon_n+\delta \epsilon/2}\frac{\sigma_t}{\Delta \epsilon}d\epsilon,
\end{equation}
where $N=\Delta\epsilon/D$ is the number of resonances in the integrated region. In the case of narrow resonances, this integral converges quickly and thus the bounds can be extended to infinity. Making the substitution $\rho = (\epsilon-\epsilon_n)/\Gamma_{\mathrm{tot}}$ we can re-write this as
\begin{align}
\label{eq:XSectionFinal}
\overline{\sigma} &\approx \frac{\pi}{D p^2}\int_{-\infty}^{\infty}\frac{\Gamma_{\mathrm{el}}}{\rho^2+1/4}d\rho\notag{} \\
&=\frac{2\pi^2 \Gamma_{\mathrm{el}}}{D p^2},
\end{align}
which is the energy-averaged optical cross-section (where we have factored out the momentum $p$ as it varies slowly in the case of narrow resonances). Importantly, the above result does not depend on the total width, and thus is not sensitive to the presence of possible inelastic channels.

However, we have made an implicit assumption in Eqn.~\eref{eq:XSectionSum}, specifically we summed over contributions from individual resonances in the averaging process. This is allowed if $\Gamma_{\mathrm{tot}}\ll D$, a rather strong condition which may in fact be relaxed slightly. This can be seen from the optical theorem,
\begin{equation}
\sigma = \frac{4\pi}{k} \mathrm{Im}f(0),
\end{equation}
the left hand side of which contains, amongst other things, interference terms whereas the right hand side only contains the sum of resonance contributions. Therefore the restriction on using this expression is that we neglect the interference terms which may have no definite sign and are suppressed by the averaging process. For the derivation and extended discussion, see \cite{FlambaumSushkov}. 

We can now apply Eqn.~\eref{eq:XSectionFinal} to our system of resonances where the elastic widths $\Gamma_{\mathrm{el}}$ are given by $\Gamma_n$ as in Eqn.~\eref{eq:GenWidth}. In the low energy and black hole limit ($ m<\epsilon_n \ll 1/R$) the resonance widths are always much smaller than the resonance spacings and so we write
\cite{LLV3}
\begin{align}
\label{eq:AXSection}
\bar{\sigma}_{a}^{\rm opt} &= \frac{2\pi^2 \Gamma_n}{D p^2}\notag{}, \\
&=\frac{4\pi x C^2  \epsilon r_{s}^{2}}{p \left(x+m^2 y/2\epsilon^2\right)}.
\end{align}
Taking the low momentum limit yields the expression
\begin{equation}
\label{eq:OptXSection}
\lim_{p\rightarrow 0} \bar{\sigma}_{a}^{\rm opt} = \frac{4\pi^2 r_{s}^{3} \epsilon^3 x}{p^2 (x+y/2)},
\end{equation}
where we have substituted back for mass using $m = \sqrt{\epsilon^2-p^2}$. Once again invoking the $r_s \rightarrow R$ limit $(x\gg y)$ we simplify the result to
\begin{equation}
\label{eq:FinalX}
\lim_{p\rightarrow 0} \bar{\sigma}_{a}^{\rm opt} \approx \frac{4\pi^2 r_{s}^{3} \epsilon^3 }{p^2 },
\end{equation}
which corresponds to the black hole cross section for $m \ne 0$~\cite{Unruh}.

Furthermore, as in \cite{ScalarRes}, one can replace the energy averaging procedure used previously with an average over the rapidly varying phase in the S-matrix. In doing so one finds
\begin{align}
\overline{S}&=\frac{1}{\pi}\int_{0}^{\pi}  \frac{1+\mathcal{R}- \epsilon r_{s}^{2} p C^2 (1-\mathcal{R})}
{1+\mathcal{R}+ \epsilon r_{s}^{2} p C^2(1-\mathcal{R})}e^{2i\delta_C} d\phi  \notag{} \\
&=\frac{1}{\pi}\left[\int_{0}^{\pi}  \frac{1+e^{2i\phi}- \epsilon r_{s}^{2} p C^2 (1-e^{2i\phi})}
{1+e^{2i\phi}+ \epsilon r_{s}^{2} p C^2(1-e^{2i\phi})}d\phi \right]e^{2i\delta_C}\notag{} \\
&=\frac{1- \epsilon r_{s}^{2} p C^2}
{1+ \epsilon r_{s}^{2} p C^2}e^{2i\delta_C}.
\end{align}
In the case where there are inelastic processes present, i.e. $|\mathcal{R}|<1$, the above approach is still valid so long as the elastic phase is rapidly varying. The S-matrix given above is precisely the result one obtains by setting $\mathcal{R}=0$ in the S-matrix, i.e. corresponds to the black hole result.
\section{Bound states}
\label{sec:BoundSubSec}
The scattering matrix may also be used to analyze the bound state ($\epsilon<m$) spectrum by returning to Eqn.~\eref{eq:ResStart}. For bound states we search for poles of the $S$-matrix for \textit{real} energy and therefore simply solve Eqn.~\eref{eq:ResStart} for $\epsilon$. Under the condition $\epsilon<m\ll1/R$, this yields the same expression for energy $(\epsilon < m)$ as given in Eqn.~\eref{eq:GenEnergy} for $\epsilon>m$. In the formalism given the ground state corresponds to $n=1$. For an alternative derivation see \cite{Gossel}.

We note that for $\epsilon<m$ both $p$ and $C^2$ have imaginary parts and therefore should be included when equating  components in Eqn.~\eref{eq:ResStart}. However, the aforementioned condition $\epsilon<m\ll1/R$ ensures that the imaginary contribution from $2\epsilon r_{s}^{2} p C^2$ is small compared to the potentially large (semi-classical) $2\pi n i$ term and may be ignored. The conditions imposed are compatible with the restriction given in Eqn.~\eref{eq:ExtCond1}. 

\begin{figure}[t]
\begin{center}
\includegraphics[width=0.48\textwidth]{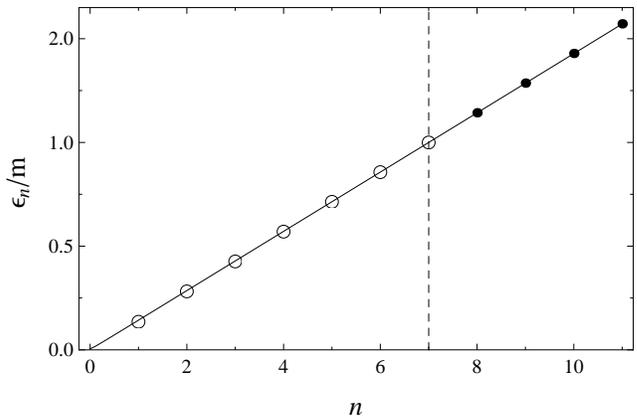}
\caption{Energies of numeric resonances (closed circles) and bound states (open circles) compared with the analytic energy (solid line) given by Eqn.~\eref{eq:GenEnergy}. The vertical dashed line represents the energy $\epsilon_n = m$. Here $r_s/R = 0.999$ and $m=0.1/R$.
\label{fig:JoinedSpectrum}}
\end{center}
\end{figure}

\begin{figure}[t!]
\begin{center}
\includegraphics[width=0.48\textwidth]{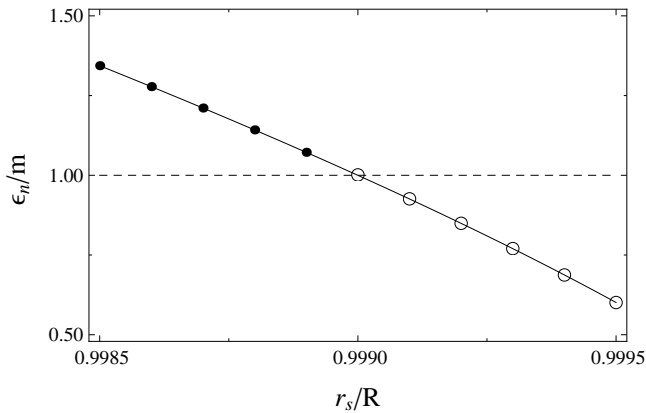}
\caption{Energies of numeric resonances (closed circles) and bound states (open circles) compared with the analytic energy (solid line) given by Eqn.~\eref{eq:GenEnergy} for the $n=7$ state. The horizontal dashed line represents the energy $\epsilon_n = m=0.1/R$. 
\label{fig:JoinedSpectrum2}}
\end{center}
\end{figure}

Equation~\eref{eq:GenEnergy} can therefore be used to describe both the continuum and bound spectra. In doing so we see that for $r_s\to R$ and a fixed value of $m$, any resonance of principal quantum number $n$ will disappear from the unbound spectrum (cross $\epsilon_n=m$ ) and be replaced by a bound state of energy $\epsilon_n$. This transition is depicted in Figures~\ref{fig:JoinedSpectrum} and~\ref{fig:JoinedSpectrum2} in the case of the Florides metric. Specifically we see how changing the value of $r_s/R$ for a fixed value of $n$ moves the state from being unbound to being bound.


\begin{figure}[t!]
\begin{center}
\includegraphics[width=0.48\textwidth]{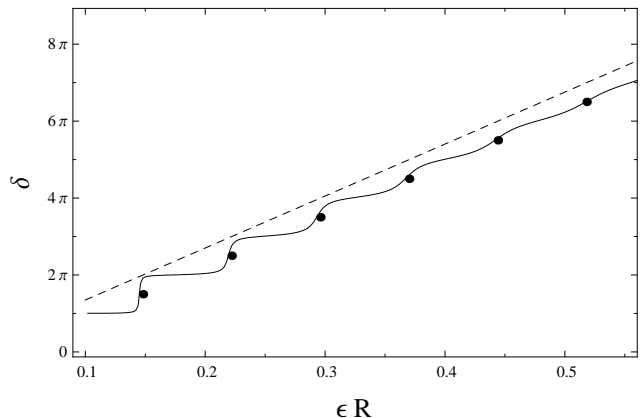}
\caption{Numerical short range phase $\delta$ (solid line) as defined in Eqn.~\eref{eq:CoulombMatch} and analytic phase (dashed line) given by Eqn.~\eref{eq:RB} in the Florides metric. The solid circles represent the analytic approximations to the energies in Eqn.~\eref{eq:GenEnergy}. Here $2\leq n \leq 7$, $r_s/R = 0.99$, and $m = 0.1/R$. This illustrates that Eqn.~\eref{eq:GenEnergy} remains a good approximation to the energies even when the resonances  begin to overlap and are no longer represented byEqn.~\eref{eq:GenWidth}.
\label{fig:PhasePlot}}
\end{center}
\end{figure}
\section{Numerical Phase Calculations}
\label{sec:NumericCalcs}
In the following sections we perform the calculations detailed prior in the Florides metric~\cite{Flor74,comment2}, characterized by
\begin{equation}
\label{eq:FInt}
e^{\nu(r)}=\frac{(1-r_s/R)^{3/2}}{\sqrt{1-r_sr^2/R^3}},\quad
e^{\lambda(r)}=\left(1-\frac{r_s r^2 }{R^3}\right)^{-1}.
\end{equation}
Calculations in this metric are performed twice: once numerically (exact solutions) and once using the semi-classical method described previously. In each case we use Eqn.~\eref{eq:KG} to generate an interior wave equation for the given metric. 

Exact solutions are obtained by numerically solving the wave equations using the boundary condition $\psi (0)=1$, $\psi '(0)=0$ with the help of {\em Mathematica} \cite{math}. This solution provides a real boundary condition for the {\em exterior} wave function at $r=R$. Equation~\eref{eq:ExteriorWave} is then integrated outwards to large distances $r\gg r_s$. As discussed prior to Eqn.~\eref{eq:Coulomb_scatter}, in this asymptotic region Eqn.~\eref{eq:ExteriorWave} takes the form of the non-relativistic Schr\"odinger equation for a particle with momentum $p$ and unit mass in the Coulomb potential with charge $Z=-r_s \left(\epsilon ^2+p^2\right)/2$. Hence, we match it with the asymptotic Coulomb solution \cite{LLV3}
\begin{equation}\label{eq:CoulombMatch}
\psi (r)\propto \sin \left[p r - (Z/p ) \ln (2 p r) +\delta_C+
\delta \right]
\end{equation}
where $\delta_C = \arg \Gamma (1+i Z/p ) $ is the Coulomb phase shift,
$\Gamma (x)$ being the Euler gamma function, and $\delta $ is the short-range
phase shift. The latter is determined almost exclusively by the interior
equation, and carries important information about the behaviour of the
wave function at $r\lesssim R$. In all cases discussed the numeric phase referred to is $\delta$ given in Eqn.~\eref{eq:CoulombMatch}. An example of this numerically calculated phase for the Florides metric is given in Fig.~\eref{fig:PhasePlot}. The body radius $R$ is set to $1$ for all numeric calculations i.e. all lengths are measured in units of $R$.

The phase possesses steps of height $\pi$ at the resonance positions $\epsilon_n$. We fit the step profile of an individual resonance to the  Breit-Wigner function
\begin{equation}
\delta(\epsilon  \simeq\epsilon_n)= \delta_n + \arctan\left[\frac{\epsilon-\epsilon_n}{\Gamma_n/2}\right]
\end{equation}
where $\delta_n$ is a constant, from which
we extract the numeric resonance widths and positions $\Gamma_n$ and $\epsilon_n$.
\begin{figure}[t!]
\begin{center}
\includegraphics[width=0.48\textwidth]{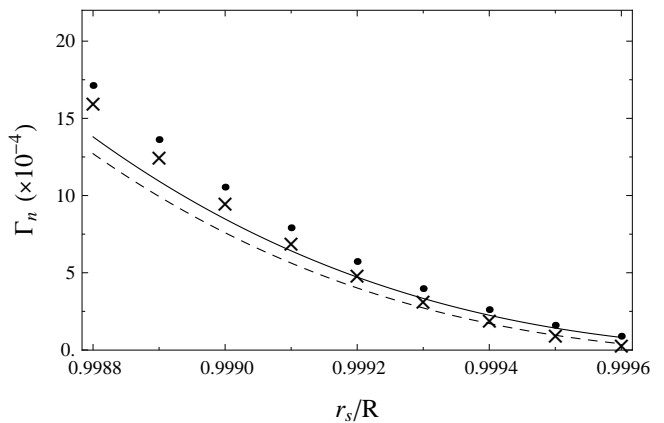}
\caption{Numeric (dots and crosses) and analytic (solid and dashed lines) widths in the Florides metric as a function of $r_s/R$. The dots and solid line correspond to $m=0$, whereas the crosses and dashed line correspond to $m=0.1/R$. In both cases the analytic widths are given by Eqn.~\eref{eq:GenWidth}
\label{fig:FloridesWidth}}
\end{center}
\end{figure}
\begin{figure}[t!]
\begin{center}
\includegraphics[width=0.48\textwidth]{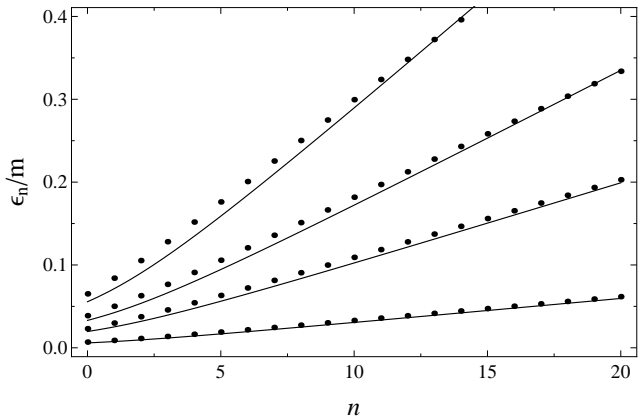}
\caption{Numeric (dots) and analytic (solid lines) energy levels in the Florides metric as a function of principal quantum number $n$ (ground state corresponds to $n=0$). Here $m = 5/R$ and $r_s =0.980$, $0.990,$ $0.995,$ $0.999$ with the shallower, lower value, lines corresponding to higher $r_s$ values.
\label{fig:FloridesLevels}}
\end{center}
\end{figure}
Using the Florides metric coefficients~\eref{eq:FInt} in Eqn.~\eref{eq:SepIntegrals} yields
\begin{align}
x &=  A (1-r_s/R)^{-3/4}-R \ln(1-r_s/R),\notag{}\\
y &= B  (1-r_s/R)^{3/4} 
\end{align}
where $A \approx 1.198R$ and $B \approx 2.622R$ (for $r_s \to R$).

Substituting these values in to Eqns.~\eref{eq:GenWidth} and ~\eref{eq:GenEnergy} we arrive at the resonance widths for the Florides case, plotted against the numerical values in Fig.~\ref{fig:FloridesWidth}. Additionally, Eqn~\eref{eq:GenEnergy} is compared against numerics for $\epsilon<m$ in Fig.~\ref{fig:FloridesLevels}.
\section{The strong energy condition and bounds on $r_s/R$}

As noted previously, much work has been done on examining the nature of the $r_s/R\to1$ limit~\cite{Anisotropy1,Anisotropy2,Anisotropy3,Anisotropy4}. These results suggest that under certain conditions, including anisotropy of the metric, one may take $r_s/R\to 1$ in a physically sensible way. However, a recent work examined these treatments in light of the strong energy condition and found that for static spherically symmetric bodies we should have $r_s/R<0.98$~\citep{0.98Ref}. While the results presented so far in this work only examine the cross section in the formal limit $r_s/R\to 1$, showing that it is equal to the pure black hole case for low $\epsilon$, we can broaden our question to ask how applicable our results are at lower $r_s/R$.
To answer this it is useful to examine the Soffel metric, an analytic continuation of the Schwarzschild \textit{interior} metric made to work beyond the Buchdahl limit. The Soffel metric~\cite{Soffel77} is characterised by
\begin{equation}
e^{\nu(r)} = \left(1-\frac{r_s}{R}\right) \exp\left[-\frac{r_s (1-r^2/R^2)}{2R (1-r_s/R)}\right]
\end{equation}
with $e^{\lambda(r)}$ as in the Florides case.

In the process of examining the effect of non-zero $m$ on the resonance structure --- the main emphasis of this work --- we performed numeric calculations on the Florides metric only, because the Soffel $dt^2$ metric coefficient is such that the effect of mass is exponentially suppressed as $r_s/R \to 1$.

\begin{figure}
\begin{center}
\includegraphics[width=0.48\textwidth]{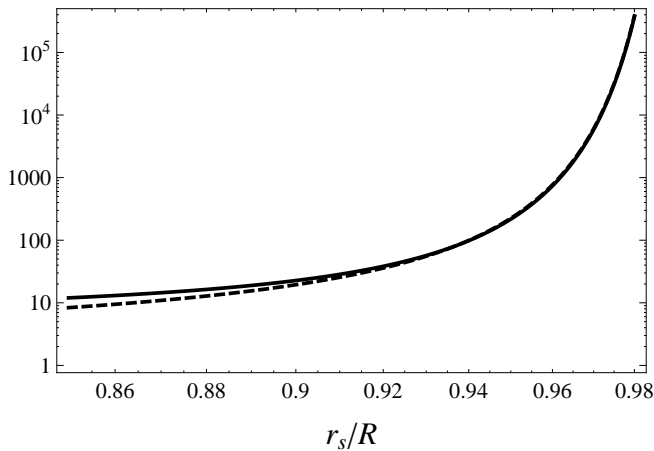}
\caption{Numeric verification of $x_{\mathrm{S}}$ in Eqn.~\eref{eq:SoffelXY}. Here we compare the analytic (solid line) and numeric (dashed line) results of the first integral in  Eqn.~\eref{eq:SepIntegrals} for the Soffel metric.
\label{fig:SoffelVerify}}
\end{center}
\end{figure}

To examine just how close one can get to the black hole case with $r_s/R<0.98$, it is useful to numerically compare the possible values of the absorption cross section. For the Soffel metric, the integrals defined following Eqn.~\eref{eq:SepIntegrals} are given by
\begin{align}
\label{eq:SoffelXY}
x_{\mathrm{S}}&= \sqrt{\frac{\pi}{r_s}} \exp\left[\frac{r_s}{4R(1-r_s/R)}\right], \notag{} \\
y_{\mathrm{S}}&= \frac{\pi}{4}\sqrt{1-r_s/R}.
\end{align}
These integrals are computed assuming $r_s/R \to 1$, but as we see in Fig.~\eref{fig:SoffelVerify}, we needn't be very close to $1$ for the approximation to be valid. Comparison of Eqns.~\eref{eq:OptXSection} and \eref{eq:FinalX} show that, for $p~\to~0$, the cross section for the near-black-hole case tends to that of the pure black hole when
\begin{equation}
\frac{x_\mathrm{S}}{x_\mathrm{S}+ m^2 y_\mathrm{S}/2\epsilon^2 } \to 1
\end{equation}
which, in light of the above expressions, clearly occurs well before $r_s/R = 0.98$. For example, it is within $10^{-5}$ of unity at $r_s/R = 0.97$ with $p = 10^{-4}$, and $m = 0.1$. One subtlety that emerges, however, is how small $p$ must be. In addition to using $m^2/\epsilon^2 \to 1$ in the above expression (a weak condition), we have made the simplification
\begin{equation}
C^2 \to \frac{\pi r_s \epsilon^2}{p}
\end{equation}
in going from Eqn.~\eref{eq:AXSection} to~\eref{eq:OptXSection}. This remains valid for $p \ll \epsilon$ (see Eqn.~\eref{eq:CFactor}) and is independent of $r_s/R$.

Coupled with the fact that the distance between the levels in the region of $\epsilon \approx m$ for $r_s/R = 0.98$ is $\sim 10^{-6}$ (much less than the value of $p$ used in the above example), we see that not only can we average over many resonances using the optical model, but that the absorption cross section is already rapidly tending to that of a pure black hole. As an example, for $p = 3 \times 10^{-6}, m = 0.1$, and $r_s/R = 0.98$, the difference between $\bar{\sigma}_{a}^{\rm opt}$~\eref{eq:AXSection} and the cross section of a black hole~\eref{eq:FinalX} is 0.005\%.

Thus, even though the $r_s/R \to 1$ limit may be disallowed under certain physical restrictions, we still find that our approach elucidates
rich and interesting physics in a sufficiently strong (but not disallowed) gravitational field.
\section{Conclusions}
By considering the scattering matrix for a scalar particle in the gravitational field of a static, spherically symmetric near-black-hole object, we determine the energy spectrum of the bound states and the unbound (scattering) resonances. In the black hole limit ($r_s\to R$) the energy density of both the bound and unbound spectra tends to infinity as both spectra collapse and become quasi-continuous. Importantly, the lifetimes of the resonant states also tend to infinity in this limit giving rise to effective absorption in a purely elastic scattering scenario. By utilising the optical model we calculate the cross section for absorption into these resonances. In the black hole limit this cross section exactly equals the cross section in the pure black hole case (for low particle momentum). Thus black-hole properties may emerge in a non-singular static metric prior to the formation of a black hole as in the $m=0$ case. Additionally, we show that both the resonance and bound state energies may be described by a single function spanning both spectra. Consequently, in the black hole limit any resonance of fixed principal quantum number $n$ will disappear and be replaced with a bound state with identical $n$. As shown previously in \cite{SchwarzsRes,Gossel}, bound states and a similarly dense spectrum of resonances also exist for massless particles in near-singular metrics with no black hole limit. We therefore conjecture that generalising to $m\neq 0$ particles in such systems would also lead to possible radiative capture, as well as lengthening of resonance lifetimes leading to a more rapid onset of the limiting behaviour.

We thank G. F. Gribakin for useful discussions. This work is supported by the Australian Research Council.


\begin{thebibliography}{99} 
\bibitem{ScalarRes} V.~V.~Flambaum,G.~H.~Gossel and G.~F.~Gribakin, Phys. Rev. D {\bf 85}, 084027
(2012).

\bibitem{DiracRes} G. H. Gossel, J. C. Berengut, and V. V. Flambaum, and G.~F.~Gribakin,
Phys. Rev. D {\bf 88}, 027501 (2013)

\bibitem{BosonRes}  Y. V. Stadnik, G.~H.~Gossel,  V.~V.~Flambaum and J. C. Berengut,
Euro. Phys. J. C {\bf 73}, 2605 (2013).

\bibitem{Gossel} G. H. Gossel, J. C. Berengut, and V. V. Flambaum,
Gen. Relativ. Gravit. {\bf 43}, 2673 (2011).

\bibitem{DiracBound} A. F. Spencer-Smith, G. H. Gossel, J. C. Berengut, and V. V. Flambaum,
Gen. Relativ. Gravit. {\bf 45}, 613 (2013).

\bibitem{Anisotropy1}R.~L.~Bowers and E.~P.~T.~Liang,
The Astrophysical Journal {\bf 188}, 657 (1974).

\bibitem{Anisotropy2} C.~G.~B\"ohmer and T.~ Harko,
Class. Quant. Grav. {\bf 23}, 6479 (2006).

\bibitem{Anisotropy3}A.~F\"uzfa, J.~M.~Gerard, and D.~Lambert,
Gen. Rel. Grav. {\bf 34}, 1411 (2002).

\bibitem{Anisotropy4} K.~Dev and M.~Gleiser, 
 Gen. Rel. Grav. {\bf 34}, 1793 (2002) 

\bibitem{Flor74}  P.~S.~Florides,
 Proc. R. Soc. Lond. A {\bf 337}, 529 (1974).

\bibitem{Soffel77}M.~Soffel, B.~M\"uller, and W.~Greiner,
J. Phys. A {\bf 10}, 551 (1977).

\bibitem{LLV3} L.~D.~Landau and E.~M.~Lifshitz, \textit{Quantum Mechanics},
3rd ed. (Butterworth-Heinemann, Oxford, 1977).

\bibitem{Unruh} W.~G.~Unruh, Phys. Rev. D {\bf 14}, 3251 (1976); Thesis,
Princeton Univ., 1971 (unpublished).

\bibitem{KerrElectron} W.~Yongjiu, T. ~Zhiming, 
Astrophys. Space Sci. {\bf 281}, 689 (2001).

\bibitem{Gaina} A.~B.~Gaina, I.~M.~Ternov, 
Izvestiya Vysshikh Uchebnykh Zavedenii, Fizika {\bf 10}, 71 (1988).

\bibitem{Grain} J.~Grain and A.~Barrau, 
Eur. Phys. J. C {\bf 53}, 641 (2008).

\bibitem{Lasenby2} A.~Lasenby, C.~Doran, J.~ Pritchard, A.~Caceres, S.~Dolan,
 Phys. Rev. D {\bf 72}, 105014 (2005).

\bibitem{Pindzola} M.~S.~Pindzola, 
J. Phys. B: At. Mol. Opt. Phys. {\bf 42}, 095202 (2009).

\bibitem{Barranco} J.~Barranco \textit{et al}, 
Phys. Rev. Lett. {\bf 109}, 081102 (2012).

\bibitem{Pravica} D.~W.~Pravica,
Proc. R. Soc. London A {\bf 445}, 3003 (1999).


\bibitem{Glampedakis} K.~Glampedakis and N.~Andersson,
Class. Quantum Grav. {\bf 20}, 3441 (2003). 

\bibitem{TwoSphere}  C.~G.~B\"ohmer and F.~Lobo,
 Int. J. Mod. Phys. D {\bf 17}, 897 (2007).  

\bibitem{commentP} Here the momentum at infinity, $p$, corresponds to the momentum in the asymptotic plane-wave solution at large distances, $\psi\sim e^{\pm ip r}$, in the absence of any long-range interaction.

\bibitem{ScalarScattering} M.~Yu.~Kuchiev and V.~V.~Flambaum,
Phys. Rev. D {\bf 70}, 044022 (2004).

\bibitem{comment2} For $r_s<2R/3$ the Florides metric may represent the gravitational field inside a cluster of particles moving in randomly oriented circular orbits for larger $R$ (Einstein cluster). In the case where $r_s > 2R/3$ the Florides metric does not correspond to any macroscopic object, but nevertheless represents a valid solution to the field equations. The `physical reasonableness' of such a solution is discussed in both N. K. Kofinti, Gen. Relativ. Gravit. {\bf 17}, 245 (1985) and L. Herrera and N. O. Santos, Phys. Rep. {\bf 286}, 53 (1997).

\bibitem{math}{\em Mathematica, Version 7.0}
(Wolfram Research, Inc., Champaign, IL, 2008).

\bibitem{FlambaumSushkov}  V. V. Flambaum and O. P. Sushkov,
 Nuc. Phys. A {\bf 412}, 13 (1984).

\bibitem{0.98Ref} H.~Andréasson,
Journal of Differential Equations {\bf 245}, 2243 (2008)


\bibitem{SchwarzsRes} V. V. Flambaum, G. H. Gossel, and G. F. Gribakin,
Phys. Rev. D, {\bf 86}, 044042 (2012)
\end{thebibliography}
\end{document}